\documentclass[twoside]{article}
\usepackage{fleqn,espcrc2}

\usepackage{graphicx}
\usepackage{amssymb}

\newcommand{\AmS}{{\protect\the\textfont2
  A\kern-.1667em\lower.5ex\hbox{M}\kern-.125emS}}

\title{Intrinsic C-Axis Tunnelling in BSCCO Crystals}

\author{C. E. Gough, P.J. Thomas, J.C. Fenton and G. Yang \address{Superconductivity Research Group, University of Birmingham, Edgbaston, Birmingham B15 2TT, United Kingdom}
        \thanks{Corresponding author email C.Gough@bham.ac.uk}}

\begin{document}
\begin{abstract}
Intrinsic c-axis tunnelling in the superconducting state has been measured in zero and finite fields in small mesa structures fabricated on the surface of 2212-BSCCO single crystals.  The temperature dependence of the zero-field critical current and quasi-particle conductance is related to microscopic d-wave models in the presence of impurity scattering. The strong field dependence of the c-axis critical current provides information on the correlation of flux pancakes across adjacent superconducting bi-layers. An instability in the IV characteristics is observed below 20K, which accounts for the apparent drop in critical current at low temperatures previously reported. 
\vspace{1pc}
\end{abstract}

\maketitle

\section{Introduction}
Measurements of c-axis tunnelling across adjacent superconducti-+ng bi-layers in highly anisotropic cuprate superconductors, such as 2212-BSCCO, provide important information about the tunnelling of superconducting pairs and thermally excited quasi-particles. In a field, the decrease in critical current can be related to the misalignment of flux pancakes across adjacent superconducting bi-layers. In this paper, we consider both the quasiparticle conductance and the field dependence of the critical current.
\section{Measurements}
Measurements have been made on small mesa structures, typically, shown schematically in the inset of figure 1, $20 \times 20 \mathrm{\mu m}$ and $\sim$10 nm in height corresponding to a linear array of intrinsic Josephson junctions \cite{kleiner}. Measurements have been made on a number of mesas on slightly underdoped $(\# 1)$, near optimally doped $(\# 2)$ and slightly overdoped $(\# 3)$ 2212-BSCCO crystals, as described along with experimental details in \cite{thomas}

In zero-field we observe the familiar multi-branched IV-characteristics first reported and explained by Kleiner \textit{et al}\cite{kleiner} and confirmed by many other groups. Yurgens \textit{et al} \cite{yurgens1} have already published measurements at relatively large fields\cite{yurgens2}, which we extend to smaller fields.
 
In our structures,  the first critical current is smaller than all subsequent transitions, indicating a slight degradation of the uppermost intrinsic junction either below the metal-HTS interface or at the milled surface between the current and voltage contacts (see ref. \cite{thomas}). When considering the temperature and field dependence of $J_{c}(T,B),$ we therefore take the largest critical current in the multibranched characteristics  as representative of the bulk properties. 

Measurements  inevitably involve a significant amount of self-heating,  $\sim I_{c} \times 20 \mathrm{mV}$ per intrinsic junction, heating the mesa above the measured base temperature.  To circumvent this problem, we determine the mesa temperature directly,  by monitoring the temperature dependent contact resistance between two separate contacts on the top of each mesa \cite{lt22} , which varies approximately as 1/T . The IV-measurements can then be corrected for the measured temperature rise, which can be as large as a few tens of degrees at the lowest temperatures for large amplitude voltage measurements. 
\section{Quasi-particle conductance}
Figure 1 shows a typical set of corrected  IV-characteristics, obtained quasistatically, for the 1st and 11th  phase-slip branches of a mesa on sample \#2. Once corrected for heating\cite{thomas}, the suitably scaled 11th branch can be superimposed on the 1st branch.  Such characteristics are dominated by quasi-particle tunnelling, since the  McCumber parameter $\beta_{c}\gg 1$. The temperature-corrected characteristics  are considerably more linear than any previously reported.  Additional short-pulse measurements confirm  heating to be the major source of non-linearity and back-bending  \cite{thomas}. Using the voltage at which the dynamic conductance becomes infinite\cite{yurgens3} is therefore unlikely to provide a meaningful measurement of the energy gap.

\begin{figure}[h]
\begin{center}
\input epsf
\epsfxsize=75mm  \epsfbox{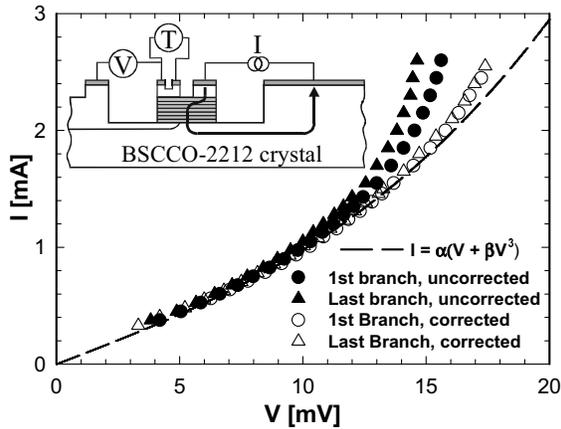} \vskip -10mm \caption{Measured IV characteristics and corrected for heating for 1st and 11th branch of \#2.} \label{fig1}
\end{center}
\end{figure}

For a d-wave superconductor with incoherent quasi-particle tunnelling, the tunnel current at low temperatures is expected to vary as $V^{3}$. In contrast,  the  conductance is almost linear over an extended energy range, with $I = \alpha V  +   \beta V^{3}$ as indicated in figure 1. Figure 2 illustrates the temperature dependence of the derived ohmic conductivity in the low-bias limit for our three samples. Also included are normal state values, extended somewhat below $T_{\mathrm{c}}$ by the application of a large field (6.6T) parallel to the c-direction. 

\begin{figure}[h]
\begin{center}
\input epsf
\epsfxsize=75mm  \epsfbox{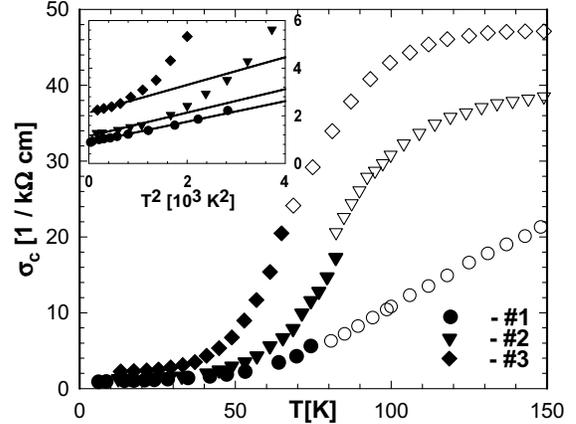} \vskip -10mm \caption{$\sigma_{c}(T)$ for mesas on three crystals}\label{fig2}
\end{center}
\end{figure}

Interestingly, there is no significant change in the  conductance nor its temperature dependence at $T_{\mathrm{c}}$. On decreasing temperature, the low-bias conductance falls continuously from its high temperature normal state value to a doping-dependent limiting value at low temperatures (see inset). 

Similar results have also been reported by Latyshev \textit{et al}\cite{latyshev} for measurements  on ion-beam milled, 2212-BSCCO,  c-axis microbridges fabricated from single-crystal whiskers. They interpret their measurements in terms of  additional impurity-induced states close to the nodes of the d-wave superconductor gap. For coherent interlayer tunnelling, such states would contribute a low bias, ohmic,  tunnel current $\sim \frac{\Gamma}{\Delta} \sigma _{n}V$ , where $\Gamma \sim \frac{h}{\tau}$ is the impurity scattering limited energy scale, in contrast to the $V^{3}$ dependence in the clean limit\cite{latyshev}. For coherent tunnelling, the conductivity would be reduced by an additional  factor $\sim \frac{\Gamma}{E_{F}}$\cite{latyshev}. We therefore expect a crossover from an  impurity-limited V-dependence to intrinsic $V^{3}$  tunnelling, when $k_{B}T^{*} \sim \Gamma$ or when $eV\sim k_{B}T^{*}$.

The only significant difference between our results and those of Latyshev \textit{et al} are the somewhat smaller values of $\sigma_{c}$ derived for samples of similar expected doping. 
\section{Field dependence of critical current}
For magnetic fields along the c-direction, the field penetrates as a series of flux pancakes with induced currents largely confined to the CuO planes. In a defect free crystal at low temperatures, the flux pancakes would align to form a uniform lattice of flux lines along the c-direction.  However, thermal fluctuations and crystal defects disturb the alignment of the pancakes, destroying the phase coherence across the planes, hence reducing the critical current by an amount $<cos\theta_{ij}>$ , where $\theta_{ij}(r)$ is the local phase difference across adjacent superconducting planes and  $<>$ represents the thermally averaged spatial average. 

Typical measurements of the field dependence of the critical current are shown in figures 3 and 4 for modest fields up to 0.1 T and for larger fields up to 7T. Measurements were taken on cooling and subsequent warming in the applied field, to ensure a nearly uniform distribution of flux. 
\begin{figure}[h]
\begin{center}
\input epsf
\epsfxsize=75mm  \epsfbox{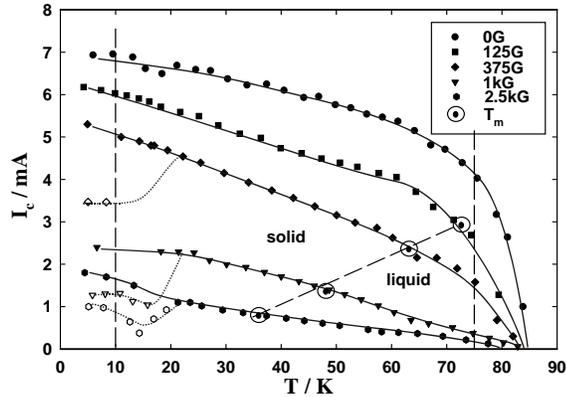} \vskip -10mm \caption{$I_{c}(T)$ at low fields for sample \#3} \label{fig3}
\end{center}
\end{figure}

\begin{figure}[h]
\begin{center}
\input epsf
\epsfxsize=75mm  \epsfbox{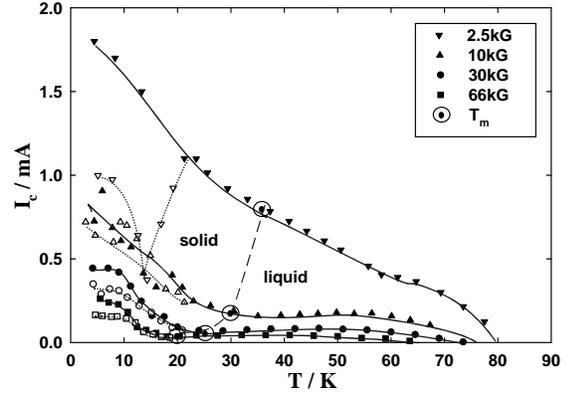} \vskip -10mm \caption{$I_{c}(T)$ at high fields for sample \#3} \label{fig4}
\end{center}
\end{figure}

In zero field, the  critical current decreases with increasing temperature, though with a different temperature dependence from BCS. The linear temperature dependence at low temperatures is inconsistent with the higher power-dependence of the c-axis penetration depth and theoretical models. This may, in part, be associated with self-heating, which was not monitored in these earlier measurements. 
 
Figures 3 and 4 suggest changes in both field and temperature dependences of the critical current as we pass from the solid to liquid vortex states, particularly at the higher fields. The position of the melting line $T(B_{m})$,  inferred from $\mu$SR and ac susceptibility measurements (see ref. \cite{lee}) on similar crystals, is indicated by the large open circles. The correlation of flux pancakes across adjacent superconducting planes therefore depends on the nature of the magnetic state.

At low temperatures $(<20\mathrm{K})$ we also observe an instability in the IV characteristics, illustrated in Figure 5.  Provided measurements are confined to the first one or two phase-slip transitions, the critical current remains high, increasing monotonically on decreasing temperature with no significant change in form of the IV-characteristics.   However, once a critical voltage has been exceeded, there is an immediate irreversible transition to an entirely new dynamic state, indicated by the lower set of characteristics in figure 5, with critical currents given by the open circles in figures 3 and 4.  The original characteristics can only be recovered by annealing the mesa above $\sim 25\mathrm{K}$. Almost identical behaviour has been observed for mesas on all three samples.  This change in phase-slip dynamics is not understood, but almost certainly accounts for the decrease in critical currents at low temperatures reported previously\cite{yurgens2}. A similar bimodal behaviour has been reported by Suzuki \textit{et al}\cite{suzuki2}.

\begin{figure}[h]
\begin{center}
\input epsf
\epsfxsize=75mm  \epsfbox{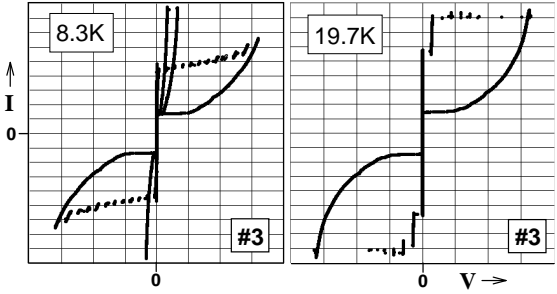} \vskip -10mm \caption{Measured characteristics illustrating bimodal behaviour at low temperatures, and stable characteristics at a higher temperature for sample \#3} \label{fig5}
\end{center}
\end{figure}

Small transverse misalignment of the pancakes vortices d ($< \gamma s$)  results in a reversed Josephson coupling energy over an area $\sim d^{2}$ per flux line,  where $\gamma$ is the anisotropy factor  and s the bilayer  spacing.  For small fields, this will  lead to a fractional decrease in critical current 
$\frac{\Delta I_{c}}{I_{c}} \sim(\frac{B}{\phi_{0}})d^{2} = \frac{B}{B^{*}}$, consistent with the low-field $e^{-\frac{B}{B^{*}}}$ dependence shown in figure 6, where $B^{*} \sim $ 725 gauss and 190 gauss for the glassy lattice state at 10K and liquid state at 75K. This corresponds to displacements d $\sim$ 165nm and $\sim$ 330nm, $\sim \gamma s$ assuming an anisotropy factor of $\sim$300.
\begin{figure}[h]
\begin{center}
\input epsf
\epsfxsize=75mm  \epsfbox{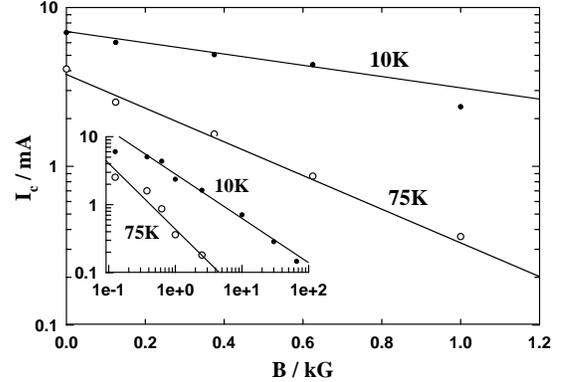} \vskip -10mm \caption{Field dependence of the critical current for sample \# 3 in the solid and liquid flux pancake states} \label{fig6}
\end{center}
\end{figure}

At higher fields, a simple model of flux pancakes of effective area $\frac{\phi_{0}}{B}$ , randomly distributed such that  $< cos  \theta_{ij}>_{\mathrm{spatial}} = 0$,   would give give a fractional reduction in Josephson current 
\begin{displaymath}
<\!\mathrm{cos}\, \theta_{ij}\, \mathrm{exp} \left( \frac{-\epsilon _{J} \phi_{0}}{B}\frac{\mathrm{cos} \,\theta_{ij}}{k_{B}T}\right) \!> \, =\frac{\epsilon_{J}}{2Bk_{B}T},
\end{displaymath}
where $ \epsilon_{J}= \frac{J_{c}\phi_{0}}{2\pi}$  is the Josephson coupling energy per unit area.  This result was derived more rigorously by Koshelev for pancakes in the liquid state \cite{koshelev1} .  A similar result is also expected for flux lines thermally diffusing in the solid \cite{koshelev2} with $J_{c}\sim \frac{1}{B^{\mu} T}$, with $\mu = 1$ for random disorder and $<1$ for increased in-plane correlation. In the solid state at low temperatures the field dependence approaches $B^{-\frac{2}{3}}$ at large fields and 1/B in the liquid state at high temperatures, as illustrated by the solid lines in Fig 5. 
\section{Acknowledgments}
This work is supported by EPSRC. We thank Peter Andrews and Gary Walsh for technical support. PJT gratefully acknowledges financial support from EPSRC.


\begin{thebibliography}{99}
\bibitem{kleiner} R. Kleiner \textit{et al}, Phys. Rev. Lett. 68 (1992) 2394
\bibitem{thomas} P.J. Thomas, J.C. Fenton, G. Yang and C.E. Gough, cond-mat/0001365
\bibitem{yurgens1} A. Yurgens \textit{et al}, Physica C 235-240 (1994) 3269
\bibitem{yurgens2} A. Yurgens \textit{et al}, Phys. Rev. B 59 (1999) 7196
\bibitem{lt22} P.J. Thomas, J.C. Fenton, G. Yang and C. E. Gough, Physica B 280, (2000) to be published
\bibitem{yurgens3} A. Yurgens \textit{et al}, Phys. Rev. B 53 (1996) R8887
\bibitem{latyshev}Yu. I. Latyshev \textit{et al}, Phys. Rev. Lett. 82 (1999) 5345
\bibitem{lee}S.L. Lee \textit{et al}, Phys. Rev. Lett. 71 (1993) 3862
\bibitem{suzuki2}M.Suzuki, T. Watanabe and A. Matsuda, Phys. Rev. Lett. 81 (1998) 4248
\bibitem{koshelev1}A.E. Koshelev, Phys.Rev.B 77 (1996) 3901 
\bibitem{koshelev2}A.E.Koshelev, Phys.Rev.B 76 (1996)1340
\end{thebibliography}
\end{document}